\documentclass[twocolumn,
superscriptaddress,
showpacs,
 amsmath,amssymb,
 aps,
 prl,
floatfix,
]{revtex4}

\usepackage{soul}
\usepackage{color} 
\usepackage{graphicx}
\usepackage{dcolumn}
\usepackage{bm}

\usepackage[T1]{fontenc}
\usepackage{lmodern}
\usepackage{color}
\usepackage{soul}

\begin{document}
\newcommand{\hlrose}[1]{\hl{#1}}
\sethlcolor{yellow}

\preprint{APS/123-QED}

\title{Spontaneous self-ordered states of vortex-antivortex pairs \\ in a Polariton Condensate}

\author{F.~Manni}%
 \email{francesco.manni@epfl.ch}
 \affiliation{%
  Institute of Condensed Matter Physics, \'{E}cole Polytechnique F\'ed\'erale de Lausanne (EPFL), CH-1015 Lausanne, Switzerland}%

\author{T.~C.~H.~Liew}
\affiliation{Division of Physics and Applied Physics, Nanyang Technological University, 63737 Singapore}

\author{K.~G.~Lagoudakis}%
\affiliation{%
  E. L. Ginzton Laboratory, Stanford University, Stanford, CA93405-4088, US}%

\author{C.~Ouellet-Plamondon}%
 \affiliation{%
  Institute of Condensed Matter Physics, \'{E}cole Polytechnique F\'ed\'erale de Lausanne (EPFL), CH-1015 Lausanne, Switzerland}%

\author{R.~Andr\'e}%
\affiliation{%
  Institut Néel, CNRS, 25 Avenue des Martyrs, 38042 Grenoble, France}%

\author{V.~Savona}
\affiliation{Institute of Theoretical Physics, Ecole Polytechnique F\'{e}d\'{e}rale de
Lausanne (EPFL), CH-1015 Lausanne, Switzerland}

\author{B.~Deveaud}%
\affiliation{%
  Institute of Condensed Matter Physics, \'{E}cole Polytechnique F\'ed\'erale de Lausanne (EPFL), CH-1015 Lausanne, Switzerland}%

\date{\today}

\begin{abstract}
Polariton condensates have proved to be model systems to investigate topological defects, as they allow for direct and non-destructive imaging of the condensate complex order parameter. The fundamental topological excitations of such systems are quantized vortices. In specific configurations, further ordering can bring the formation of vortex lattices. In this work we demonstrate the spontaneous formation of ordered vortical states, consisting in geometrically self-arranged vortex-antivortex pairs. A mean-field generalized Gross-Pitaevskii model reproduces and supports the physics of the observed phenomenology.
\end{abstract}

\maketitle


Quantized vortices are fundamental and ubiquitous entities across physics, playing a central role in mechanisms ranging from galaxy formation to phase conformation in microscopic quantum systems. They represent topological excitations of quantum degenerate Bose gases, such as Bose-Einstein condensates (BEC), superfluids and superconductors~\cite{fetter_vortices_2010}. Throughout the last decades, these systems have offered the unprecedented opportunity of studying such topological defects and their phenomenology in a direct and controlled way~\cite{fetter_vortices_2010}. Under peculiar conditions, quantized vortices have the unique property of arranging themselves in geometrically ordered structures, such as the Abrikosov lattices~\cite{abrikosov_magnetic_1957,abrikosov_nobel_2004}. Vortex lattices were first observed in type-II superconductors under magnetic fields~\cite{essmann_direct_1967}, then in both superfluids~\cite{yarmchuk_observation_1979} and atom BEC~\cite{matthews_vortices_1999,madison_vortex_2000,aboshaeer_observation_2001,schweikhard_vortex_2004}, by setting the system into rotation, and also in optical nonlinear systems~\cite{swartzlander_optical_1992,fleischer_observation_2003}. Ultimately, in the limit of high vortex density, these lattices are predicted to undergo a quantum phase transition to strongly correlated states, similar to quantum Hall states, that still represent an open experimental challenge~\cite{fetter_vortices_2010}.

Recently, exciton-polaritons have established themselves as a model two-dimensional Bose-gas~\cite{carusotto_quantum_2012}. Such quasi-particles arise as the eigenmodes of the strong-coupling regime between light and matter, which was demonstrated in planar semiconductor microcavities~\cite{weisbuch_polariton_1992}. The polariton system being dissipative in nature, due to the finite lifetime of the quasi-particles, its phenomenology is intrinsically and strongly out-of-equilibrium. Thus, continuous optical pumping is required to replenish the polariton population. Shaping of the excitation conditions allows manipulation of the condensate phenomenology in a simple way, unveiling striking physical effects. Moreover, thanks to the mixed light-matter components of polaritons, the emitted photons inherit all the properties of the quantum fluid that can, thus, be fully characterized by optical measurement of the extracavity field~\cite{savona_exact_1995}.

After the recent demonstration of polariton condensation~\cite{kasprzak_bose-einstein_2006} and superfluidity~\cite{amo_collective_2009,amo_superfluidity_2009} in semiconductor microcavities, large efforts have been devoted to the study of quantum turbulence and vorticity, under different excitation conditions, in such out-of-equilibrium quantum fluids~\cite{krizhanovskii_effect_2010,grosso_soliton_2011,sanvitto_persistent_2010} In particular, quantized vortices were demonstrated to spontaneously occur in the system as topological defects pinned by the disorder potential~\cite{lagoudakis_quantized_2008}, which naturally results from the sample growth process. In analogy with the striking demonstrations of collective vortex phenomena in atom BEC, a number of proposals and theoretical investigations considered means for the formation of vortex lattices in polariton condensates, both in the scalar~\cite{keeling_spontaneous_2008,liew_generation_2008,keeling_controllable_2010,gorbach_vortex_2010} and spinor condensate case~\cite{liew_generation_2008,keeling_controllable_2010}. Methods for detecting the rotating lattice have also been proposed for the experimental observation~\cite{borgh_robustness_2012}. Recent demonstration of vortex lattices were provided by exploiting periodic metallic structures that impose the lattice unity cell geometry~\cite{kim_dynamical_2011} and by engineering the interference of multiple independent condensates ~\cite{tosi_geometrically_2012}, bringing the topic to the forefront of polariton research.

In this letter, we demonstrate the spontaneous occurrence of polariton condensed states composed of geometrically self-arranged vortex-antivortex pairs. Each vortex having its counter-rotating partner, the overall state carries no orbital angular momentum. We perform interferometric measurements allowing the determination of the amplitude and the phase of the condensate order parameter, directly imaging the vortical entities. Theoretically, a mean-field approach reproduces the vortex-antivortex lattice formation.

The sample is the same CdTe planar semiconductor microcavity, featuring $26~\rm{meV}$ of Rabi splitting, used in our previous experiments~\cite{kasprzak_bose-einstein_2006,lagoudakis_quantized_2008,manni_spontaneous_2011}. The sample is kept in a He-cooled cryostat at approximately $4~\rm{K}$. As mentioned, a key feature for the stabilization of vortex-antivortex lattices lies in the careful shaping of the intensity profile of the pump laser.
\begin{figure}[tb]
\includegraphics[width=0.45\textwidth]{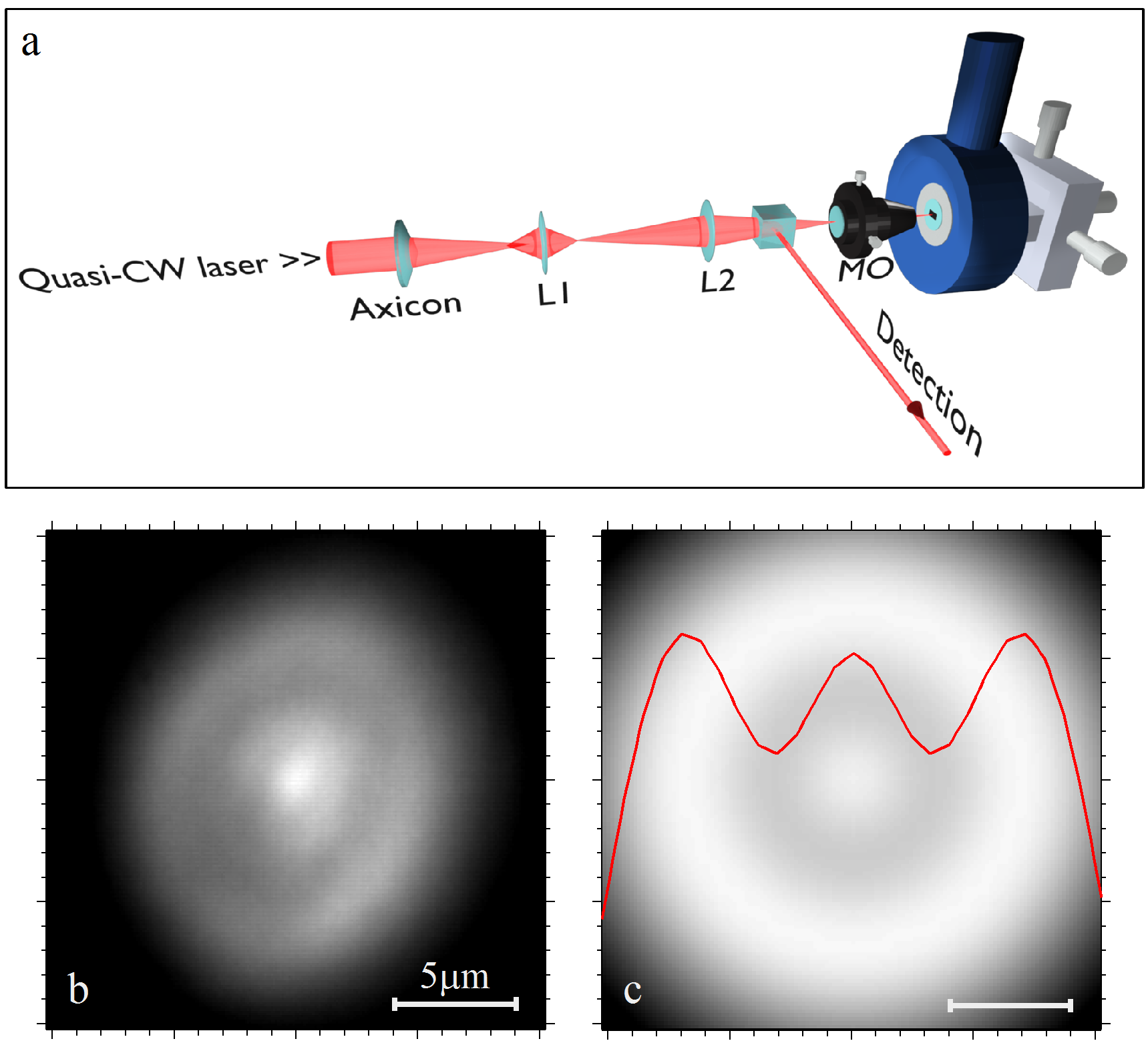}
\caption{\label{fig:figure1} (a) Scheme of the setup used to engineer the excitation. A typical shaped intensity profile is shown in (b), averaged over many disorder realizations. Intensity profile for the pump beam used in the theoretical calculations (c). The red line depicts the cross-section intensity profile of the beam.}
\end{figure}
The principle is to create a ring-shaped intensity profile, similar to a Bessel beam, with a large central spot of the same intensity as the outer ring. The reduced intensity valley between the central spot and the outer ring acts as the guide for the stabilization of vortex-antivortex pairs. Simply using a Gaussian shaped pump would not be enough, since vortices are known to be dynamically unstable in such a configuration~\cite{ostrovskaya_dissipative_2012}. The shaping of the beam is done by employing a single conical lens - $axicon$ - in combination with two lenses (L1, L2) and a $0.5~\rm{N.A.}$ microscope objective (MO), according to the scheme depicted in Fig.~\ref{fig:figure1}(a). The initial top-hat laser beam is transformed into a Bessel beam by using the axicon. In the transformation, intermediate intensity profiles are generated, as the one used in this work~[Fig.\ref{fig:figure1}(b)], which is then imaged by lenses L1 and L2 in the real space of the sample, through the MO. Fine tuning of the excitation is allowed by a system of relative micrometric positioning between the axicon and the lenses.
\begin{figure}[tb]
\includegraphics[width=0.45\textwidth]{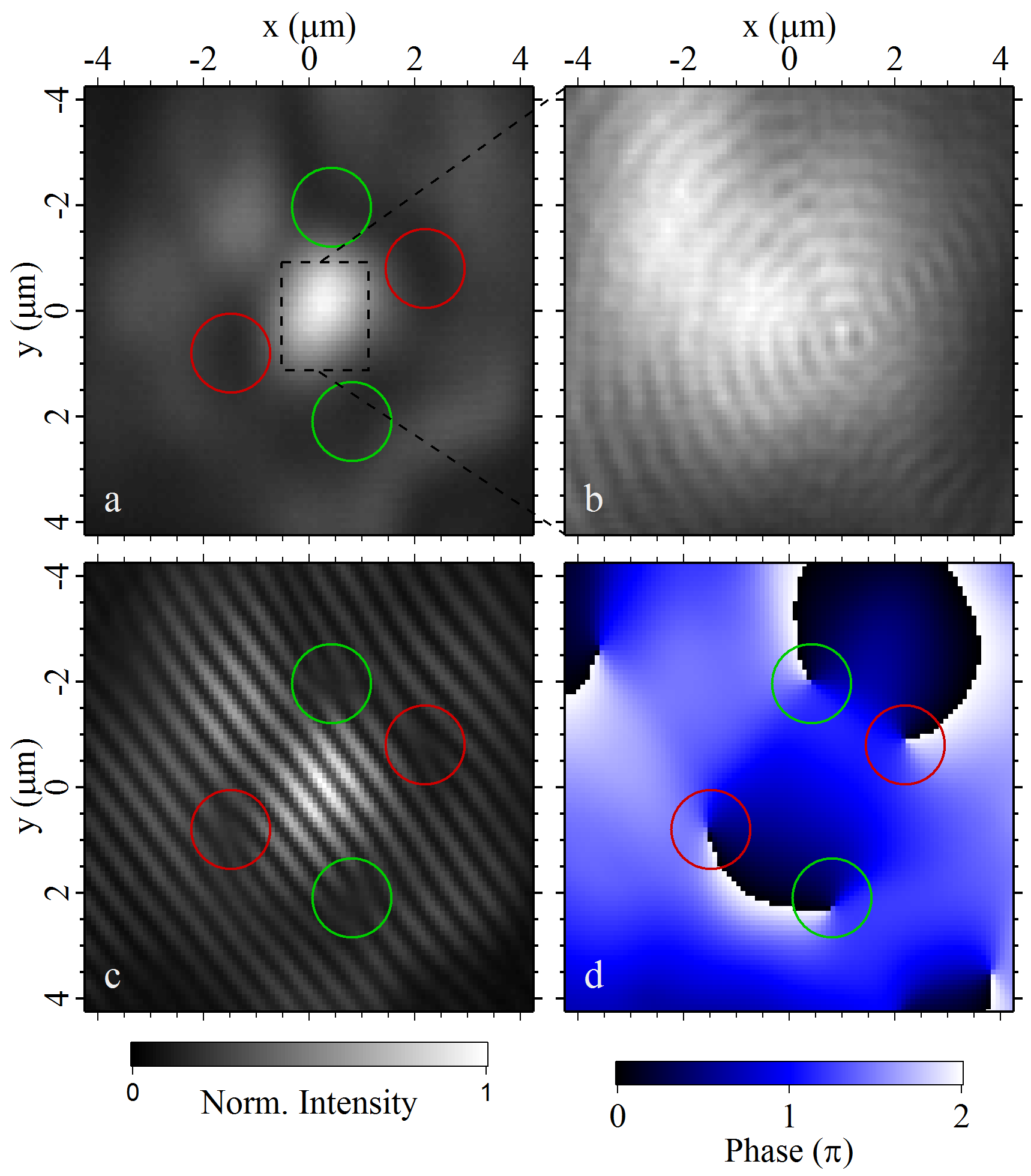}
\caption{\label{fig:figure2} (a) Polariton condensate density at a position where a four-vortex lattice is observed. (b) Magnified central part of the condensate [marked by the dashed rectangle in panel (a)] used to overlap with the full condensate to obtain the interference pattern, shown in (c). The corresponding phase structure is plotted in (d). Red (green) circles mark the position of vortex (antivortex).}
\end{figure}

We excite the system above condensation threshold with the intensity-shaped pump beam at a power of $\approx250 \rm{\mu W/cm}^{2}$ The photoluminescence (PL) coming from the sample is collected by the same MO and is then sent to a modified Mach-Zehnder interferometer, which takes the condensate image and interferes it with a 4x magnified replica the central spot (marked by the dashed rectangle in Fig.~\ref{fig:figure2}(a)). This allows to take a small part of the condensate, acting as phase reference, overlap it with the whole condensate image and extract its phase structure, as performed in Ref.~\cite{manni_spontaneous_2011}. From the interference pattern, the phase can then be extracted by digital off-axis holographic techniques~\cite{lagoudakis_observation_2009}. Despite the presence of a disorder potential~\cite{krizhanovskii_coexisting_2009}, some regions of shallow disorder are present in the CdTe sample that allow the formation of vortex lattices.

\begin{figure}[tb]
\includegraphics[width=0.45\textwidth]{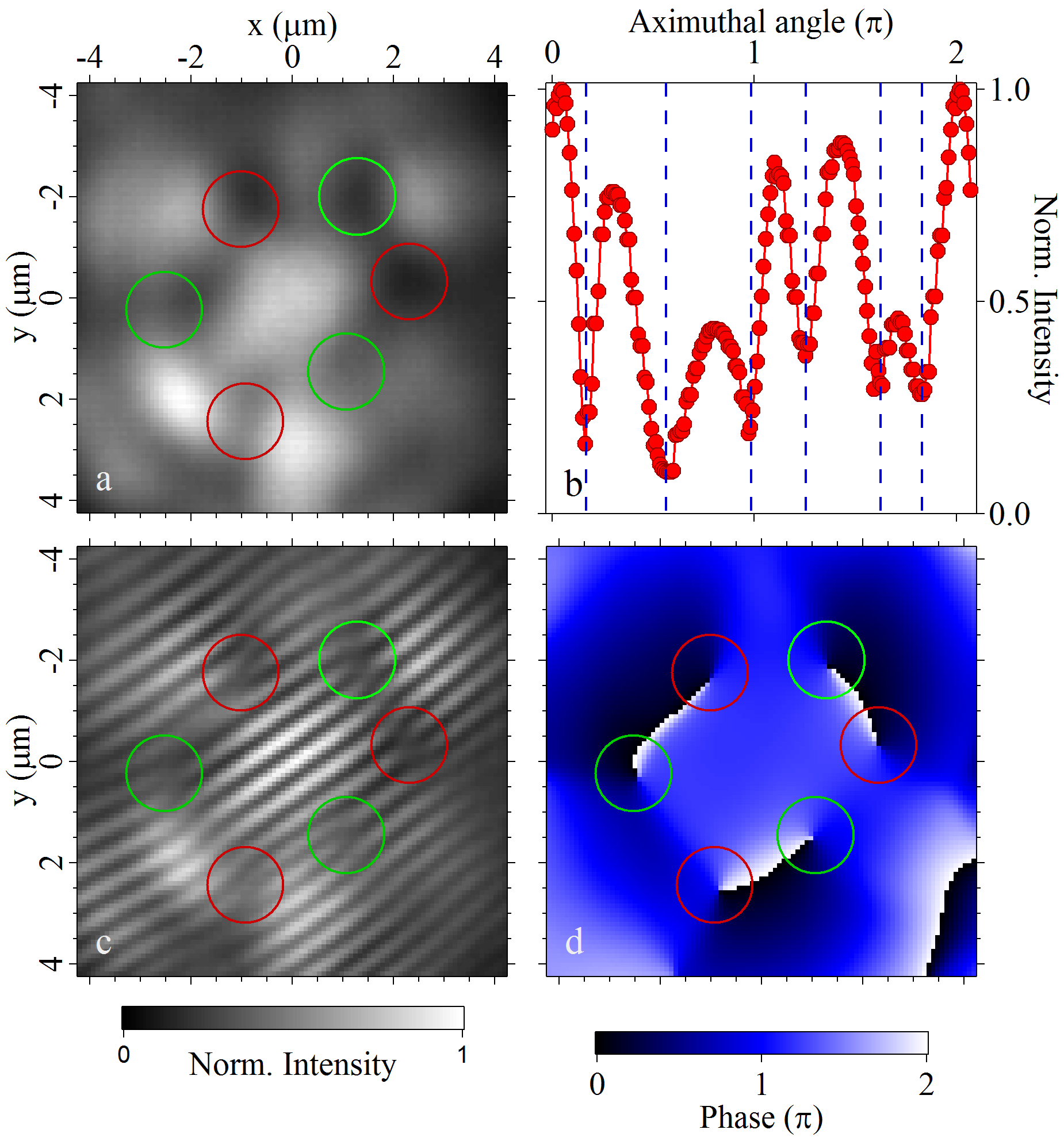}
\caption{\label{fig:figure3} (a) Polariton condensate density at a position where a six-vortex lattice is observed. (b) Profile taken around the central spot of the condensate that highlights the relative minima, identifying the core of the six vortices (each vertical line marks the minimum position). (c) Interference pattern from which the phase structure is extracted, as shown in (d). Red (green) circles mark the position of vortex (antivortex).}
\end{figure}
To explain the whole process, we first consider a set of data showing a four-vortex lattice, composed of 2 vortex-antivortex pairs. The condensate density image (Fig.~\ref{fig:figure2}(a)) is interfered with the enlarged version of the central spot of the condensate (Fig.~\ref{fig:figure2}(b)). The resulting pattern is shown in Fig.~\ref{fig:figure2}(c) together with the corresponding phase structure reported in Fig.~\ref{fig:figure2}(d). In the figure, two vortex-antivortex pairs can easily be identified, marked by the red (vortex, clockwise phase winding) and green (antivortex, counterclockwise phase winding) circles at the position of the phase dislocation. Observing the real space phase map, when circumventing the center of the condensate cloud, one finds alternately vortex/antivortex entities. Let us point out that some spurious phase singularities (with no circular marker on top) appear in a region of negligible density, outside the condensed polariton gas. They are most probably resulting from overlap of the decaying outer tails of the condensate with the edges of the flat phase reference.

It is possible to create vortex lattices of higher order, where more vortex-antivortex pairs are able to accommodate in the same alternate ordered way. By slightly increasing the size of the pump spot (in the one micron range) while keeping its intensity profile as in Fig.~\ref{fig:figure1}(b), we could find other positions on the sample where the disorder allowed the formation of the six-vortex lattice. The results are summarized in Fig.~\ref{fig:figure3}. The condensate density, featuring six clear density minima is shown in Fig.~\ref{fig:figure3}(a).The vortex cores are evidenced also by taking a circular profile of the density~\cite{data_treatment_details}, shown in Fig.~\ref{fig:figure3}(b). The vertical dashed lines identify the position of six clear relative density minima, vortex cores, along the azimuthal profile. By analysis of the orientation of the fork-like dislocations in the interference pattern [Fig.~\ref{fig:figure3}(c)] and corresponding phase structure [Fig.~\ref{fig:figure3}(d)], one can identify even better the position of the phase dislocations and the winding sign of each vortex. The red and green color circles, as in the case of Fig.~\ref{fig:figure2}, mark the alternating vortex-antivortex lattice. Despite the presence of the underlying disorder potential felt by the polaritons, the six-lattice nicely matches an hexagonal structure. Indeed, the disorder accounts for the distortions with respect to an exact symmetric, theoretically expected, Abrikosov vortex-lattice observed in other systems~\cite{yarmchuk_observation_1979,matthews_vortices_1999,madison_vortex_2000,aboshaeer_observation_2001,schweikhard_vortex_2004,swartzlander_optical_1992,fleischer_observation_2003}.

\begin{figure}[tb]
\includegraphics[width=0.48\textwidth]{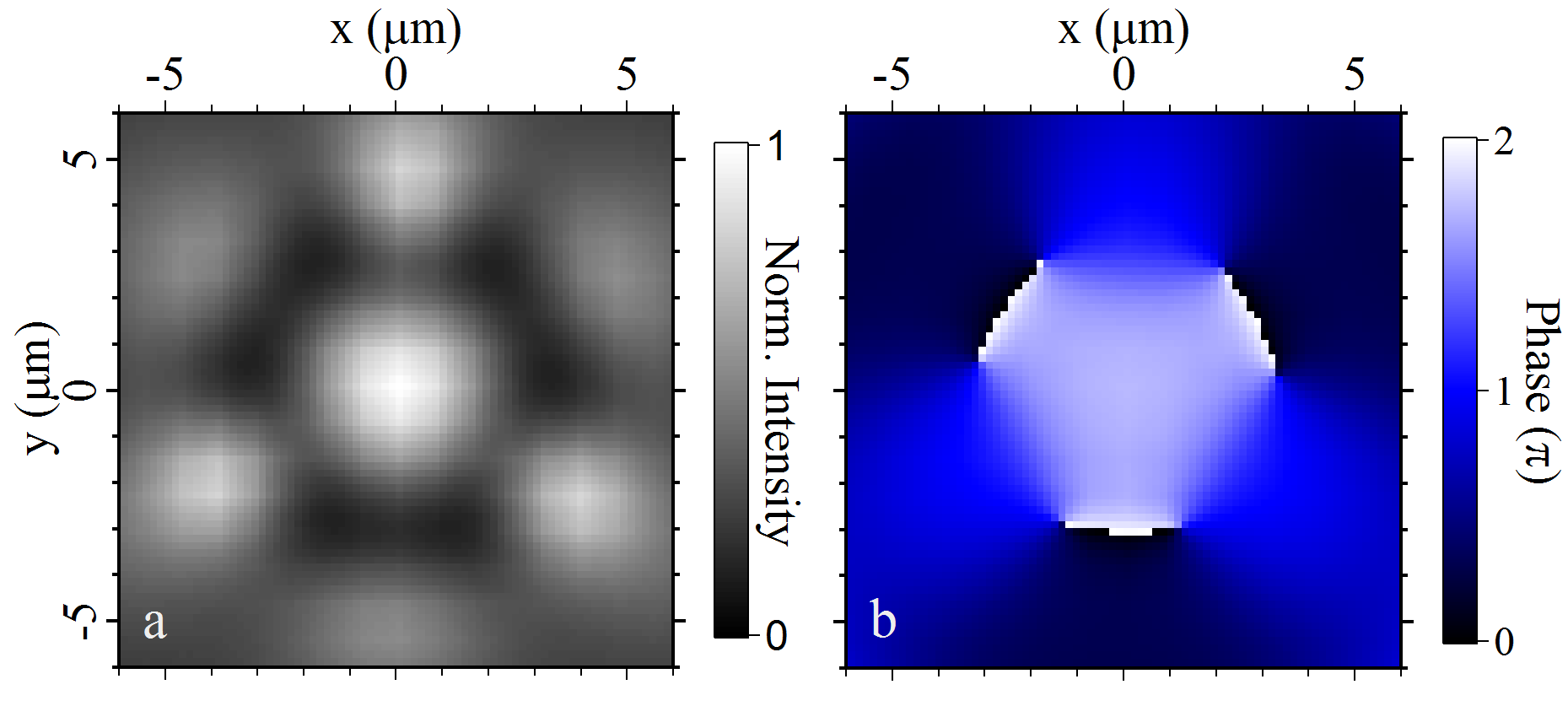}
\caption{\label{fig:figure4} Theoretically calculated polariton condensate density (a) and phase structure (b) that nicely matches the experimental results of Fig.~\ref{fig:figure3}.}
\end{figure}
Polariton condensation under non-resonant pumping is a highly non-equilibrium process. Being coherent, the condensate itself can be treated with a Gross-Pitaevskii type equation coupled to a classical rate equation for the dynamics of an exciton reservoir~\cite{wouters_spatial_2008}, which appropriately models the incoherent excitations generated by non-resonant pumping that feed and interact with the condensate. Both spectral and spatial features coming from experiments have been successfully reproduced within this theoretical framework~\cite{roumpos_gain_2010,wouters_energy_2010,christmann_polariton_2012} and also adapted to take into account the dynamic properties of polariton condensates~\cite{nardin_dynamics_2009,lagoudakis_coherent_2010,lagoudakis_probing_2011} and of spinor condensates~\cite{manni_dissociation_2012,kammann_nonlinear_2012}. The equations read:
\begin{align}
i\hbar \frac{\partial \psi(\mathbf{r},t)}{\partial t} & =\left[\hat{E}_{LP} + V(\mathbf{r}) + \frac{i\hbar}{2}\left(R_{R}n(\mathbf{r},t) - \gamma_{c}\right)\right] \psi(\mathbf{r},t) \label{eq:schrodinger} \\
\frac{\partial n(\mathbf{r},t)}{\partial t} & = -\left(\gamma_{R} + R_{R}|\psi(\mathbf{r},t)|^{2}\right) n(\mathbf{r},t) + P(\mathbf{r}) \label{eq:reservoir}
\end{align}%
where $\psi(\mathbf{r},t)$ is the polariton condensate in the mean-field representation and $n(\mathbf{r},t)$ is the intensity distribution of the incoherent excitonic reservoir, directly injected by the non-resonant pump. For the purpose of this work the polarization degree of freedom of polaritons is neglected. $\hat{E}_{LP}$ is the polariton kinetic energy operator, accounting for the non-parabolicity of the dispersion of lower branch polaritons. $V(\mathbf{r})$ represents an effective potential, given by the mean-field shift induced by polariton-polariton interactions ($g$), the interaction of polaritons with the reservoir ($g_{R}$) and an additional pump induced shift (G)~\cite{wouters_spatial_2008}; $V(\mathbf{r}) = \hbar g|\psi(\mathbf{r},t)|^{2} + \hbar g_{R}n(\mathbf{r},t) + \hbar GP(\mathbf{r})$, where $g$, $g_{R}$ and $G$ are constants. $P(\mathbf{r})$ represents the spatial pump distribution, as in Fig.~\ref{fig:figure1}(c). The intra-condensate interactions between polaritons being present but small, the dominant effects determining the induced effective potential come from the effect of the exciton reservoir on condensed polaritons~\cite{roumpos_gain_2010}. $\gamma_{c}$ and $\gamma_{R}$ represent the decay rates of condensed polaritons and reservoir excitons, respectively. $R_{R}$ is the stimulated scattering rate of excitons from the reservoir to the condensate and thus represents the condensation rate in the system.
The system of two coupled equations~\ref{eq:schrodinger} and~\ref{eq:reservoir} is numerically solved in time. The initial conditions for seeding the condensate consist of an initial random low-intensity white noise. Theoretically, the system reaches the same steady vortex lattice pattern independent of the initial noise. The initial noise may however affect the time required for formation of the lattice. One can note that the vortex lattice pattern is not cylindrically symmetric (although it has a rotational symmetry of third order). Theoretically, the breaking of cylindrical symmetry could be caused by the initial noise, in which case the pattern would form with a different overall orientation (but otherwise identical shape) from shot to shot. Our system is however not perfectly symmetrical to begin with; imperfections in the pump profile and disorder already break the system symmetry allowing the pattern to form with the same orientation from shot to shot and be observed in multi-shot averaged experiments~\cite{manni_spontaneous_2011}.

Figure~\ref{fig:figure4} shows the intensity and phase of the polariton condensate steady state, calculated with this method~\cite{simulation_param} under the custom intensity-shaped pump beam of Fig.~\ref{fig:figure1}(c). A good agreement is found between theory and experiment (Fig.~\ref{fig:figure4}) both in the condensate density and the phase structure, which feature the three vortex-antivortex pairs. The excitation laser, with its reduced intensity central ring, acts as a \emph{guide} and as a trap for the vortices and favors the formation of spatially coherent states with regular lattice of vortex-antivortex pairs within the \emph{trapping} potential due to its shape.

The number of vortices composing the lattice is partly sensitive to the profile of the pump, in terms of its size and shape. The use of larger sized pumps allows for larger vortex lattices with more vortex-antivortex pairs. Furthermore, it is worth mentioning that the disorder plays an important role in the determination of the energy of the condensate, substantially influencing the potential landscape felt by the particles, so that only at a few specific sample positions could we obtain the engineered pump-induced trapping of the vortices and observe the lattice formation. At such positions, slight adjustment of the pump beam alignment allows fine tuning of the symmetry and stability of the lattice itself.

The theoretical results presented in Fig.~\ref{fig:figure4} show that a steady-state has formed in the system for the chosen form of excitation. However, it should be appreciated that steady-state solutions may not exist for all pump profiles and intensities. For a smaller sized-pump we have also obtained periodic solutions of a rotating vortex lattice, where the number of vortices and antivortices need not be an even number. The experimental observation of such states is a challenging task, either requiring single-shot time-resolved interferometry or by means of more advanced interferometric techniques~\cite{borgh_robustness_2012}.

In this work we have reported on the formation and stabilization of vortex lattices, composed of vortex-antivortex pairs, in a polariton condensate. Ordered patterns are observed under proper intensity-shaped non-resonant optical pumping. The shaping is such to create a trap for vortices that are seen to arrange in a geometrical vortex lattice. A delicate interplay between the excitation shape and the underlying disorder potential pins the orientation of the spatial patterns, allowing their detection in time-integrated experiments. We make use of a mean-field approach in the form of a generalized Gross-Pitaevskii model that is able to reproduce the lattice formation under the same kind of excitation conditions used in the experiments. The control over the lattice formation could prove to be useful in Bose gases when considering the few particle per vortex limit, which is expected to give rise to quantum phase transitions to highly correlated states~\cite{fetter_vortices_2010} -- a world yet to be explored. Recent long-lifetime polaritons achieved in GaAs cavities~\cite{nelsen_coherent_2012} represent to this extent really promising systems.

The authors would like to thank M. Borgh and J. Keeling for fruitful discussions. This work was supported by the Swiss National Science Foundation through NCCR ``Quantum Photonics'' and SNSF project 135003.

\bibliography{VortexLattices}

\begin{thebibliography}{44}
\bibitem[{dat()}]{data_treatment_details}
\bibinfo{note}{The circular profile is obtained from the interference pattern,
  from which both the the condensate density and phase are extracted.}
\bibitem[{sim()}]{simulation_param}
\bibinfo{note}{We used the following parameters in the simulations:
  $\hbar\gamma_{c} = 1$ meV, $\hbar\gamma_{R} = 10$ meV, $\hbar R_{R} = 0.1$
  meV$\mu m^{2}$, $G = 0.01 \,\mu m^{2}$, $\hbar g_{R} = 0.033$ meV$\mu
  m^{2}$.}
\end{thebibliography}

\end{document}